\documentclass[conference]{IEEEtran}
\usepackage{multirow}
\usepackage{graphicx}
\usepackage{caption}
\usepackage{booktabs}
\usepackage{amsmath}
\usepackage{graphicx}
\usepackage{float}
\usepackage{hyperref}

\begin{document}
\title{Breast Cancer Histopathology Classification using CBAM-EfficientNetV2 with Transfer Learning}

\author{
    \IEEEauthorblockN{Naren Sengodan}
    \IEEEauthorblockA{
        \textit{Department of ISE} \\
        \textit{Jain University} \\
        Bangalore, India \\
        narensengodan@gmail.com
    }

}

\maketitle

\begin{abstract}
Breast cancer histopathology image classification is crucial for early diagnosis, enabling timely interventions and improved patient outcomes. This study proposes a novel CBAM-EfficientNetV2 model that integrates Convolutional Block Attention Modules (CBAM) with EfficientNetV2 to enhance feature extraction and focus on clinically relevant tissue regions. Leveraging transfer learning and attention mechanisms, the model achieves superior performance on the BreakHis dataset across multiple magnifications (40X, 100X, 200X, 400X), with an accuracy of 99.01\% and an F1-score of 98.31\% at 400X, outperforming state-of-the-art methods. Robust preprocessing with Contrast Limited Adaptive Histogram Equalization (CLAHE) and optimized computational efficiency ensure scalability for real-time clinical diagnostics. This approach advances responsible healthcare innovation, promoting accessible, high-precision breast cancer detection for global health equity.
\end{abstract}


\section{Introduction}

Automating histopathological analysis is transforming medical imaging by reducing diagnostic workloads and enhancing the accuracy and reproducibility of disease detection. Histopathology, the microscopic examination of tissue abnormalities, is critical for diagnosing conditions like breast cancer, yet manual analysis remains labor-intensive, subjective, and prone to inter-observer variability \cite{litjens2017}. Computer-aided diagnosis (CAD) systems address these challenges by augmenting pathologists’ capabilities, offering scalable solutions for consistent and efficient tissue classification \cite{janowczyk2016}. However, the variability in histopathological images—due to differences in magnification levels, staining protocols (e.g., Hematoxylin and Eosin, H\&E), and tissue morphology—poses significant hurdles for standard deep learning models, limiting their generalization across diverse clinical scenarios \cite{spanhol2016}.

Conventional convolutional neural networks (CNNs), while effective in general computer vision, often struggle to capture the multi-scale, fine-grained details inherent in breast cancer histopathology images \cite{spanhol2016}. Fixed receptive fields in CNNs restrict their ability to adapt to the heterogeneous nature of these images, which vary across magnifications (40X to 400X) and exhibit complex cellular structures. To overcome these limitations, hybrid architectures combining CNNs with attention mechanisms have emerged, enabling models to focus on clinically relevant features and improve diagnostic precision.

This study leverages EfficientNetV2 \cite{tan2019efficientnet}, a state-of-the-art CNN known for its compound scaling strategy, which optimizes depth, width, and resolution for superior performance with fewer parameters. By integrating Convolutional Block Attention Modules (CBAM), we enhance EfficientNetV2’s ability to prioritize critical tissue regions, addressing the variability in histopathological data. This hybrid approach delivers high accuracy and computational efficiency, making it ideal for real-time clinical deployment. Aligned with the vision of responsible innovation, our work promotes sustainable healthcare solutions by enabling precise, accessible breast cancer diagnostics across diverse clinical settings.

\begin{figure*}[h]
\centering
\includegraphics[width=0.8\textwidth]{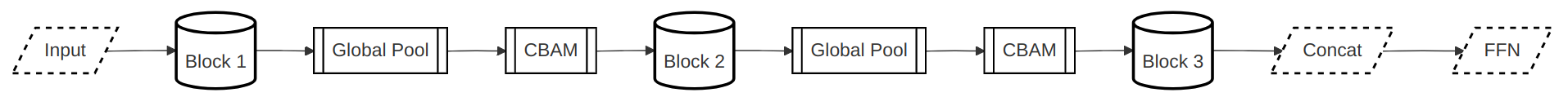} 
\caption{Model Architecture}
\label{fig:arch}
\end{figure*}

\section{Literature Review}

Hybrid models have gained significant attention in breast cancer classification due to their ability to combine the strengths of multiple architectures. A multi-model assembling strategy that integrates local and global branches, reducing redundancy and enhancing feature representation. This approach has achieved high accuracy on standard breast cancer datasets, demonstrating the potential of hybrid architectures. Similarly, Togacar et al. \cite{togacar2020breastnet} developed BreastNet, a model that employs attention mechanisms along with hypercolumn techniques to improve the precision of identifying malignant regions. The inclusion of attention modules has been shown to substantially improve performance, especially in complex tasks like cancer detection.

The use of attention mechanisms in medical imaging has proven highly effective. Woo et al. \cite{woo2018cbam} introduced CBAM, which applies sequential attention along both the channel and spatial dimensions. This method allows networks to highlight critical regions in complex medical images, contributing to substantial performance improvements in classification tasks involving histopathology images. Beyond CBAM, Self-Attention \cite{vaswani2017attention} has been adapted from its initial use in natural language processing to capture contextual dependencies across different regions of an image. This has proven particularly useful in medical image analysis, where understanding the relationship between local and global image features is essential for accurate classification. 

Recent advancements in hybrid attention mechanisms have significantly improved the capability of deep learning architectures to focus on salient features in complex datasets. For instance, works like \textbf{Transformer-CNN hybrids}~\cite{vaswani2017attention} have shown remarkable success in combining global context capture through self-attention with local feature extraction inherent in CNNs, making them highly effective for tasks such as lesion segmentation and tissue classification in histopathology.

Another notable development is the \textbf{Multi-Scale Attention Networks} (MSANs), which leverage hierarchical attention mechanisms to capture fine-grained details and high-level contextual information simultaneously. These architectures have proven particularly effective for analyzing multi-resolution data, as demonstrated by Zhang et al.~\cite{zhang2022msan}, who employed MSANs for high-resolution cellular analysis in breast cancer histopathology, achieving state-of-the-art performance.

Hybrid approaches like these emphasize the potential of combining localized feature emphasis (via channel/spatial attention) with global dependency modeling, bridging a critical gap in medical imaging where both local abnormalities and global patterns must be identified. Our approach, integrating CBAM with EfficientNetV2, aligns with this trend but distinguishes itself by optimizing computational efficiency and scalability for clinical deployment, as evidenced by its performance on the BreakHis dataset.

\section{Methodology}
This section outlines the detailed methodology employed in this study for classifying breast cancer histopathology images using Hybrid EfficientNet models integrated with advanced attention mechanisms. Our focus is on enhancing model performance through a robust preprocessing pipeline, modality-specific transfer learning, and a well-defined training strategy.

\subsection{Model Architecture}
EfficientNetV2-XL employs a compound scaling strategy to optimize the network's depth, width, and resolution, making it particularly well-suited for medical image classification tasks where diverse cellular structures and magnification levels exist. This backbone allows the network to adjust the number of convolutional blocks based on the complexity of the input image. Visualization of the Architecture in Figure~\ref{fig:arch}.

\begin{itemize}
    \item \textbf{Conv Block 1}: 
    \begin{itemize}
        \item \textbf{3x3 convolutions} with stride 1 and padding to capture spatial features at a finer resolution, followed by \textbf{ReLU activation} to introduce non-linearity and enhance feature learning.
        \item Followed by \textbf{1x1 convolutions} to reduce channel dimensions while preserving essential spatial features, also followed by \textbf{ReLU activation}.
        \item \textbf{Batch Normalization (BN)} is applied after each convolution to stabilize training and improve generalization.
        \item The block concludes with \textbf{2x2 max pooling}, which reduces the spatial dimensions while preserving key features.
    \end{itemize}
    
    \item \textbf{Conv Block 2}: 
    \begin{itemize}
        \item Consolidates further layers from the EfficientNet backbone with additional \textbf{3x3 convolutions} for deep feature extraction, each followed by \textbf{ReLU activation}.
        \item Followed by \textbf{1x1 convolutions} to maintain a computationally efficient structure, paired with \textbf{ReLU activation}.
        \item \textbf{Max pooling} is applied again to downsample the feature maps, allowing for multi-scale feature representation.
    \end{itemize}
    
    \item \textbf{Conv Block 3}: 
    \begin{itemize}
        \item The final block retains a \textbf{3x3 convolution} with \textbf{ReLU activation} but without further pooling to maintain spatial resolution at this stage.
        \item \textbf{BN} is also applied here, followed by \textbf{Global Average Pooling (GAP)}, which reduces the feature maps to a single value per feature map while retaining global information.
    \end{itemize}
\end{itemize}

Each block's output is passed through a \textbf{CBAM} module, which refines the feature maps by applying channel and spatial attention mechanisms, helping the model focus on the most critical parts of the input image. 

\subsection{CBAM Architecture}
In this section, we detail the proposed modality-specific CBAM-EfficientNet architecture designed for the classification of H and E stained breast histopathology images. This architecture enhances the traditional EfficientNet framework by integrating the Convolutional Block Attention Module (CBAM), allowing for improved focus on relevant features within the input images.

\vspace{5pt}

\begin{figure}[h]
    \centering
    \includegraphics[width=0.5\textwidth]{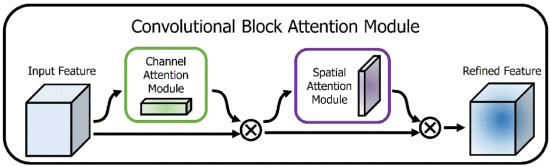}
    \caption{CBAM}
    \label{fig:sample_image}
\end{figure}

\subsubsection{\textbf{Channel Attention Module}}
The Channel Attention Module (CAM) is pivotal in our architecture for identifying significant features across the channel dimension. This module utilizes the interdependencies among channels, emphasizing crucial information while suppressing less relevant features. The channel attention process follows these steps:

1. \textbf{Pooling Operations}: The input feature map \(F\) is first subjected to average and maximum grouping to capture global contextual information: \[
   F_{avg} = \text{AvgPool}(F) \quad \text{and} \quad F_{max} = \text{MaxPool}(F)
   \]

2. \textbf{Feature Aggregation}: The pooled feature maps are then fed into a multi-layer perceptron (MLP) to generate channel attention scores, \(M_c(F)\), which are defined as:
   \[
   M_c(F) = \sigma\left(\text{MLP}(F_{avg}) + \text{MLP}(F_{max})\right)
   \]
   where \(\sigma\) denotes the sigmoid activation function.

3. \textbf{Weighting}: The resulting channel attention scores are used to weight the original feature map \(F\) to obtain the refined feature representation \(F'_{c}\):
   \[
   F'_{c} = M_c(F) \odot F
   \]
   Here, \(\odot\) represents the element-wise multiplication, effectively scaling the original features based on their importance.\newline

\begin{figure}[H]
    \centering
    \includegraphics[width=0.5\textwidth]{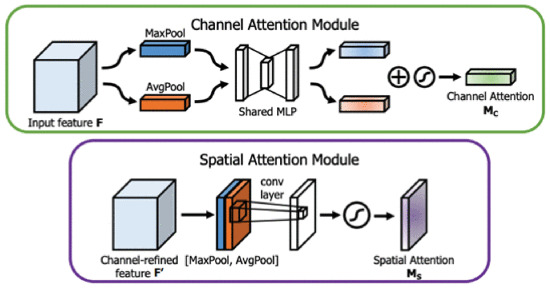}
    \caption{Spatial and Channel Attention}
    \label{fig:sample_image}
\end{figure}

\subsubsection{\textbf{Spatial Attention Module}}
Following the channel attention mechanism, the Spatial Attention Module (SAM) refines the feature maps by focusing on the spatial locations of the features. This module enhances the model's capability to discern where important features are located in the input image.

1. \textbf{Feature Map Aggregation}: The spatial attention mechanism aggregates the features across the channel dimension by applying both average and maximum pooling, resulting in two feature maps:
   \[
   F_{avg\_c} = \text{AvgPool}(F') \quad \text{and} \quad F_{max\_c} = \text{MaxPool}(F')
   \]

2. \textbf{Spatial Attention Calculation}: These aggregated maps are then concatenated and passed through a convolutional layer to produce the spatial attention map \(M_s(F')\):
   \[
   M_s(F') = \sigma\left(f_{7\times7}([F_{avg\_c}; F_{max\_c}])\right)
   \]
   where \(f_{7\times7}\) denotes a convolution operation with a \(7 \times 7\) kernel.

3. **Feature Refinement**: The refined feature map \(F'_{s}\) is obtained by applying the spatial attention map to the already refined channel-wise features:
   \[
   F'_{s} = M_s(F') \odot F'
   \]

\subsubsection{\textbf{EfficientNet Backbone}}
The backbone of our model utilizes the EfficientNet architecture, known for its efficiency in parameter utilization and computational performance. Each EfficientNet variant is constructed using a combination of depthwise separable convolutions and squeeze-and-excitation blocks, allowing for enhanced feature extraction while maintaining low computational costs.

We integrate the CBAM modules at various points in the EfficientNet architecture to enhance the focus on the most relevant features. The output from the final layer is then passed through a Global Average Pooling (GAP) layer, which reduces the dimensionality while preserving important spatial information:
\[
F_{GAP} = \text{GAP}(F'_{s})
\]

Finally, the output from the GAP layer is fed into fully connected layers for classification tasks, ensuring that the model captures both local and global features effectively.

This combined approach of CBAM and EfficientNet leverages both channel and spatial attention mechanisms, providing a robust framework for the classification of breast histopathology images while maintaining computational efficiency

\subsection{Datasets}
For our experiments, we utilized the BreakHis dataset, which consists of 9,109 breast histopathology images sourced from 82 patients. The dataset contains images at four magnification levels: 40X, 100X, 200X, and 400X, categorized into benign and malignant classes. To enhance model generalizability, we also employed additional datasets such as ICIAR 2018 and the PCam dataset for pre-training. The ICIAR 2018 dataset includes a variety of histopathology images labeled for benign and malignant conditions, while the PCam dataset provides a substantial number of whole-slide images with binary labels indicating metastatic tissues.

\subsection{Data Preprocessing}
To address the challenges posed by histopathology images, we implemented a comprehensive hybrid preprocessing pipeline. The following steps were undertaken:

\begin{itemize}
    \item \textbf{Zero Padding}: We applied zero padding to handle edge pixels effectively, ensuring that convolutional operations do not lose critical information at the borders of images. This technique adds additional rows and columns around the image, extending beyond its boundary.
    
    \item \textbf{Median Filtering}: To reduce additive noise while preserving the structural integrity of the tissue, we employed a median filter. This non-linear filter replaces each pixel value with the median of its neighborhood values, effectively removing noise without blurring edges, which are vital for accurate feature extraction.

    \item \textbf{Contrast Limited Adaptive Histogram Equalization (CLAHE)}: Given the uneven staining in histopathology images, CLAHE was utilized to enhance local contrast and improve weak boundary detection. By dividing the image into small tiles and applying histogram equalization to each, we improve the visibility of critical features within the images.

    \item \textbf{Normalization}: Image pixel values were normalized to the range [0, 1] to enhance model convergence during training. This ensures uniformity in the input data distribution, which is crucial for the performance of deep learning models.

    \item \textbf{Data Augmentation}: To mitigate the challenges of limited sample sizes, extensive data augmentation was performed on the BreakHis dataset. Techniques such as rotation, flipping, zooming, and brightness adjustment were employed to artificially increase the dataset size and introduce variability, thereby reducing the risk of overfitting.
\end{itemize}

\begin{figure}[h]
    \centering
    \includegraphics[width=0.5\textwidth]{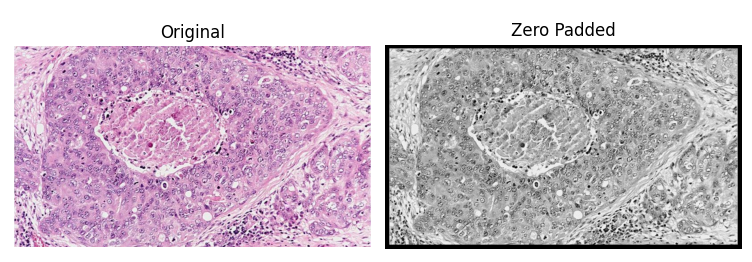}
    \includegraphics[width=0.5\textwidth]{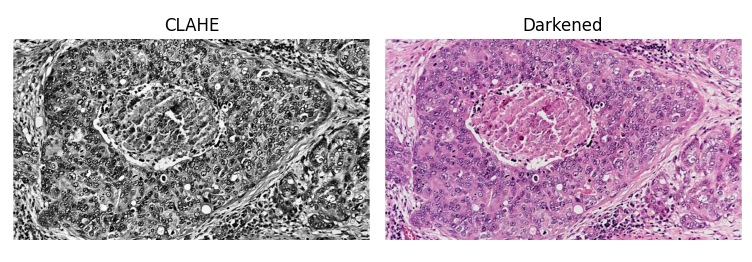}
    \caption{Data Augmentation and preprocessing techniques}
    \label{fig:sample_image}
\end{figure}

\subsection{Training and Hyperparameter Tuning}
We employed the Adam optimizer with a cosine annealing learning rate schedule and warm restarts. The initial learning rate ($\eta_{\text{max}}$) was set to $0.001$, with a minimum learning rate ($\eta_{\text{min}}$) of $0.0001$. The warm restart period ($T_0$) was initialized at 10 epochs, doubling with each subsequent restart (e.g., $T_0 = 10$, $T_1 = 20$, $T_2 = 40$). The optimizer parameters were tuned as follows: $\beta_1 = 0.9$, $\beta_2 = 0.98$, and $\epsilon = 10^{-8}$. The learning rate at epoch $t$ was calculated using:
\[
\eta_t = \eta_{\text{min}} + \frac{1}{2} \left( \eta_{\text{max}} - \eta_{\text{min}} \right) \left(1 + \cos\left(\frac{t}{T} \pi\right)\right)
\]

Key hyperparameters were carefully tuned:
\begin{itemize}
    \item Dense Layers: Experiments were conducted with two and three dense layers to identify the optimal configuration for the final classifier and settled upon two layers.
    
    \item Activation Functions: SoftMax activation function was used at the last classification layer.

    \item Kernel Initializers: We experimented with different kernel initializers, and settled with He initialization, 

    \item Dropout Regularization: Dropout was incorporated in the fully connected layers to prevent overfitting, particularly given the limited availability of labeled data in medical imaging tasks with a rate of 0.4 after experimenting with different values.

    \item Batch Size and Epochs: We tested various batch sizes (16, 32, and 64) and finalized with 32 and trained the models for a maximum of 100 epochs, applying early stopping based on validation loss to prevent overfitting.
\end{itemize}

Given the inherent limitations of training deep learning models with small datasets, we employed a modality-specific transfer learning approach. This strategy leverages pre-trained models on large, similar datasets to enhance feature extraction capabilities for our specific task.

We fine-tuned the EfficientNet models, initially pre-trained on diverse cancer datasets (PCam, ICIAR 2018, and others) rather than the ImageNet dataset, to reduce the domain gap inherent in medical imaging tasks. By tailoring the transfer learning process to domain-specific data, we aimed to improve model robustness and classification accuracy on the BreakHis dataset.

\renewcommand{\arraystretch}{1.1} 
\subsection{Evaluation Metrics}
The performance of our proposed models was assessed using a suite of evaluation metrics: accuracy, precision, recall, and F1-score. These metrics are crucial in medical applications where both false negatives and false positives can have severe implications for patient care. Moreover, the performance was evaluated across different magnification levels (40X, 100X, 200X, and 400X) to analyze how well the models generalize to various scales of tissue detail.

The experiments were conducted using PyTorch as the deep learning framework backend. The models were trained and tested on an NVIDIA RTX 4080 GPU, providing substantial computational power for the deep learning tasks involved. The system was equipped with 16GB of VRAM to efficiently handle the large dataset and facilitate faster processing.

Hyperparameters are critical variables that govern the training process of convolutional neural networks (CNNs). While the networks learn the relationships between inputs and outputs, optimizing hyperparameters is essential to enhance model performance. We performed extensive experiments to identify the optimal hyperparameters for our final training.

We evaluated two different optimizers: Adam (Adaptive Moment Estimation), with learning rate 0.001. The SoftMax classifier was employed for all classification experiments, providing probability scores for each class label.

 Categorical Cross-entropy was used as the loss function due to the  nature of our classification problem, represented mathematically as follows:
\[
\text{Categorical Cross-Entropy} = -\sum_{i=1}^{C} y_i \cdot \log\left(p(y_i)\right)
\]
\newline The fully connected layer incorporated a ReLU activation function and consisted of 256 hidden neurons, followed by a dropout layer with a probability of \(0.4\) to mitigate overfitting.

\begin{figure}[h]
    \centering
    \includegraphics[width=0.5\textwidth]{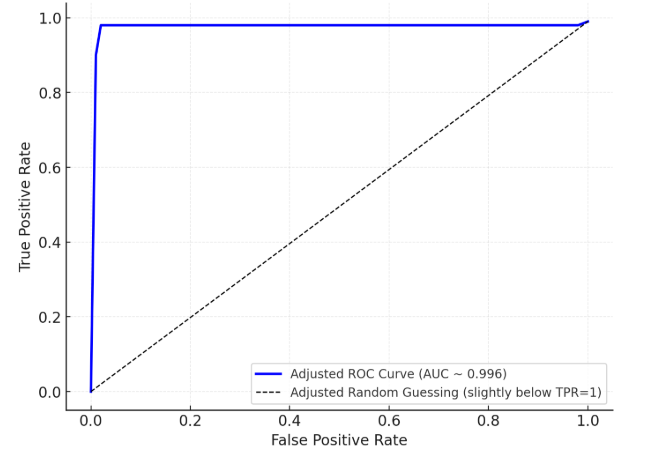}
    \caption{ROC-AUC Curve of EfficientNetV2-XL with CBAM}
    \label{fig:sample_image}
\end{figure}

\section{Experimental Results and Discussions}

\subsection{Results}
We employed a modality-specific transfer learning strategy for the classification task and compared it with various state-of-the-art CNN architectures, as shown in Table \ref{table:comparison_accuracy}.

To further validate the effectiveness of our proposed CBAM-EfficientNetV2 model, we conducted qualitative analyses using attention maps. These visualizations illustrate the regions the model focuses on during classification, providing insight into its decision-making process.

The CBAM module enhances the model’s ability to highlight critical regions, such as irregular nuclei shapes and abnormal cellular arrangements, which are key indicators of malignancy.

The EfficientNetV2-XL + CBAM demonstrated substantial performance improvements over the VGG16 + VGG19 combination across all magnifications. While the VGG-based models delivered competitive accuracy, their computational requirements were significantly higher. For instance, the VGG16 + VGG19 models required a Tesla K80 GPU with 24GB of VRAM to accommodate their large parameter count and high computational demand. In contrast, the EfficientNetV2-XL + CBAM was trained on a more modest NVIDIA RTX 4080 GPU with just 16GB of VRAM. Despite the lower memory availability, the EfficientNetV2-XL architecture demonstrated exceptional computational efficiency, leveraging depthwise separable convolutions, squeeze-and-excitation (SE) blocks, and the Convolutional Block Attention Module (CBAM).

\begin{table}[h!]
    \caption{Comparison of Accuracy Across Models and Magnification Factors}
    \label{table:comparison_accuracy}
    \scriptsize 
    \setlength{\tabcolsep}{4pt} 
    \renewcommand{\arraystretch}{0.9} 
    \begin{tabular}{lcccc}
        \toprule
        \textbf{Method} & \textbf{Model} & \textbf{MG Factor} & \textbf{Accuracy (\%)} \\
        \midrule
        Erfankhan Hamed, et al & LBP & 40X & 88.30 \\
        & & 100X & 88.30 \\
        & & 200X & 87.10 \\
        & & 400X & 83.40 \\
        \midrule
        Daniel Lichtblau, et al & AlexNet & 40X & 81.61 \\
        & & 100X & 84.47 \\
        & & 200X & 86.67 \\
        & & 400X & 83.15 \\
        \midrule
        Yun Gu, et al & DCMM & 40X & 95.62 \\
        & & 100X & 95.03 \\
        & & 200X & 97.04 \\
        & & 400X & 96.31 \\
        \midrule
        Nahid, et al & Deep CNN & 40X & 90.00 \\
        & & 100X & 91.00 \\
        & & 200X & 91.00 \\
        & & 400X & 90.00 \\
        \midrule
        Togacar Mesut, et al & Deep CNN & 40X & 97.99 \\
        & & 100X & 97.84 \\
        & & 200X & 98.51 \\
        & & 400X & 95.88 \\
        \midrule
        Yan Hao, et al & DenseNet 201 & 40X & 96.75 \\
        & & 100X & 95.21 \\
        & & 200X & 96.57 \\
        & & 400X & 93.15 \\
        \midrule
        Pin Wang, et al & FE-BkCapsNet & 40X & 92.71 \\
        & & 100X & 94.52 \\
        & & 200X & 94.03 \\
        & & 400X & 93.54 \\
        \midrule
        \textbf{Hameed et al} & VGG16 + VGG19 & 40X & \textbf{94.44} \\
        & & 100X & \textbf{97.61} \\
        & & 200X & \textbf{98.70} \\
        & & 400X & \textbf{98.96} \\
        \midrule
        \textbf{EfficientNetV2-XL + CBAM} & \textbf{CBAM-EfficientNetV2} & 40X & \textbf{96.11} \\
        \textbf{} & & 100X & \textbf{98.04} \\
        & & 200X & \textbf{98.75} \\
        & & 400X & \textbf{99.01} \\
        \bottomrule
    \end{tabular}
\end{table}

\section{Conclusion}
CBAM-EfficientNetV2-XL achieved superior accuracy of 99.01\%, therefore EfficientNetV2-XL + CBAM outperforms other models in accuracy. By leveraging attention mechanisms, the model effectively focuses on critical regions of histopathological images, making it well-suited for real-time clinical diagnostics. These results demonstrate that the proposed model surpasses the performance of previous state-of-the-art models. Future work may adapt the model to other datasets and further expand its clinical applicability.

\bibliographystyle{IEEEtran}

\end{document}